\documentclass[12pt,preprint]{aastex}
\def\ltorder{\mathrel{\raise.3ex\hbox{$<$}\mkern-14mu
             \lower0.6ex\hbox{$\sim$}}}
\def\day{{}$^{\rm d}$\llap{.}}
\def\msun{{M_\odot}}

\def\deg{\mathrel{\raise.9ex\hbox{$\circ$}\mkern-7mu
             \lower0.4ex\hbox{$\cdot$}}}

\begin{document}

\title{The Distance of the First Overtone RR Lyrae Variables
in the MACHO LMC Database: A New
Method to Correct for the Effects of Crowding}

\author{C. M. Clement, X. Xu and A. V. Muzzin}
\affil{Department of Astronomy and Astrophysics, 
University of Toronto, Toronto, ON, Canada, M5S 3H8}

\begin{abstract}

Previous studies have indicated that many of the
RR Lyrae variables in the LMC have properties similar to the ones in
the Galactic globular cluster M3.
Assuming that the M3 RR Lyrae  variables 
follow the same relationships among period, temperature,
amplitude and Fourier phase parameter $\phi_{31}$ as their
LMC counterparts, we have used the
M3 $\phi _{31}-\log P$ relation to identify the
M3-like unevolved first overtone RR Lyrae  variables in 16 
MACHO fields near the LMC bar.
The temperatures of these variables were calculated from the
M3 $\log P - \log T_{eff}$ relation 
so that the extinction could be derived for each star separately. 
Since blended stars have lower amplitudes for a given period, 
the period-amplitude relation should be a useful tool for
determining which stars are affected by crowding. 
We find that the low amplitude LMC RR1 stars are brighter than 
the stars that fit the M3 period-amplitude relation and
we estimate that at least 40\% of the stars are blended. 
Simulated data for three of the crowded stars illustrate that an
unresolved companion with $V \sim 20.5$ could account for the
observed amplitude and magnitude.
We derive a corrected mean apparent magnitude
$\langle V_0 \rangle =19.01 \pm 0.10$ (extinction) $\pm 0.02$
(calibration) for the 51 uncrowded 
unevolved M3-like
RR1 variables.  Assuming that the unevolved RR1 variables
in M3 have a mean absolute magnitude $M_V=0.52 \pm 0.02$
leads to an LMC distance modulus $\mu = 18.49 \pm 0.11$ 
\end{abstract}


\section{Introduction}
\label{sec-introduction}

Variable stars  are useful standard candles for determining  
the distances to nearby galaxies, but
crowding can be a major source of uncertainty in the measurement of
these distances. 
If the image of a variable is blended with that of another
star, its apparent magnitude will be too bright. As a result,
its derived distance will be too
small. It is not always possible to recognize which stars 
are affected by the crowding
so  the problem is usually addressed by taking a statistical 
approach. 
Another consequence of image blending is that
the amplitude of light variation is reduced. 
Thus, if one can determine which
variables have low amplitudes for their periods, the blended
stars can be flagged.
In this paper, we propose a method, based on the period-amplitude
relation, for identifying crowded stars in the LMC. 
We test our method on the 330
RR Lyrae stars in the MACHO database that we classified as bona fide
RR1\footnote{We use the system of notation that Alcock et al. (2000b) 
introduced for RR Lyrae variables: RR0 for fundamental mode, RR1 for
first-overtone and RR01 for double-mode (fundamental and first-overtone),
instead of the traditional RRab, RRc and RRd.} variables in our previous
study (Clement et al. 2002; Alcock et al. 2004, hereafter referred to
as A04).

Investigations of the LMC RR Lyrae variables 
by the MACHO collaboration have indicated
that many of them have characteristics similar to the ones in the
Galactic globular cluster M3. In a preliminary study that included
500 RR Lyrae stars, Alcock et al. (1996) calculated that the  mean period of 
the RR0 
variables with a $V$ amplitude of $0.8$ mag was 0\day 552,
compared with 0\day 480 for M107,
0\day 507 for M4, 0\day 543 for M3 and M72 and 0\day 617
for M15. In a subsequent paper, (Alcock et al. 2000a),
a more rigorous selection that included only the least 
crowded RR0 stars\footnote{The
percentage flux inside the point-spread function box contributed by
neighboring stars was estimated and only the $\sim 20\%$ least
crowded stars were included in their sample.} in 16 fields near the bar
was made.
The  ridgeline of the period-amplitude relation they plotted
for the RR0 variables 
was similar to the M3 ridgeline based on the data of Kaluzny et al.
(1998).
The population of double-mode variables in the LMC is another
feature that indicates a similarity to M3.
More than 150 of the 181 RR01 stars discovered by
Alcock et al. (1997, 2000b) have fundamental mode
periods between 0\day 46 and 0\day 50. The only globular
clusters known to have RR01 variables with periods
in this range are M3 (Corwin et al. 1999, Clementini et al. 2004)
and IC 4499 (Clement et al. 1985, Walker \& Nemec 1996).

\begin{figure}
\includegraphics{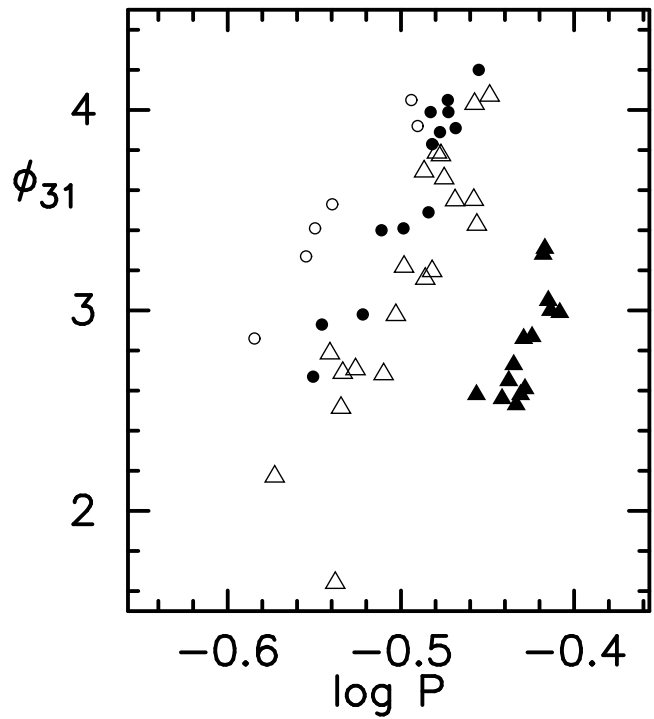}
\caption{ Plot of $\phi_{31}$ vs. $\log P$ for the RR1
variables in four well studied Galactic globular clusters:
the Oosterhoff type I clusters
M107 (open circles), M5 (solid circles) and M3 (open triangles)
and the Oosterhoff type II cluster M68 (solid triangles).
The $\phi_{31}$ values  plotted here 
for these four clusters
were determined by Clement \& Shelton (1997), Kaluzny et al. 
(2000), Cacciari et al (2005) and 
Walker (1994) respectively.
Their metal abundances are $-1.10$, $-1.32$,
$-1.50$ and $-2.43$ on the $\rm{Fe}_{\rm{II}}$ metallicity
scale of Kraft \& Ivans (2003).
\label{fig-phigc}}
\end{figure}

In this study, we use Simon's (Simon \& Lee 1981)
Fourier decomposition technique. 
Previous investigations have pointed to a relationship among
his Fourier phase parameter $\phi_{31}$, and the 
luminosity and metal abundance.
Evidence for this can be seen in
Figure \ref{fig-phigc} which shows 
$\phi_{31}-\log P$ plots for the RR1 variables in
four well studied globular clusters: 
M107, M5, M3 and M68. 
The diagram illustrates 
that the RR1 variables in a given cluster show
a sequence of $\phi_{31}$ increasing with period and that the lower the
cluster metallicity, the more the sequence is shifted 
to longer periods in the plot. Since 
the luminosities of RR Lyrae variables are known to be
correlated with 
metal abundance, a plot like
Figure \ref{fig-phigc} should be useful for identifying
a group of RR1 stars with similar luminosities.\footnote{Simon 
\& Clement (1993) derived equations relating
masses, luminosities and temperatures to pulsation period
and $\phi_{31}$ based on  hydrodynamic
pulsation models. 
Later Catelan (2004) and Cacciari et al. (2005, 
hereafter referred to as CCC) both demonstrated that the
equations should not be applied to individual stars, but CCC
also noted that Fourier parameters could be used
for estimating the average luminosity of a group of
stars, after careful and proper calibration. In this investigation,
we do not use the Fourier parameters to determine physical properties
for individual stars.}
The  $\phi_{31}-\log P$ plot that A04 made for
the LMC 
indicated that most of the RR1 variables are similar to
the ones in M2, M3 and  M5. 
We therefore assume that the luminosities of the LMC RR1 variables
are comparable to the ones in these three clusters.
Since M3 is the most variable-rich globular cluster 
and also because CCC recently made a detailed multicolor
and Fourier study of its RR Lyrae variables,
we will compare the LMC stars with the ones in M3.
Our modus operandi will be to use $\phi_{31}$ to
identify the M3-like RR1 variables
in the LMC, then use the period-amplitude relation to
select the ones that are uncrowded and
apply the M3 distance modulus to derive
their absolute magnitudes.
The $\phi_{31}$ parameter is effective for this analysis because
it is not altered by crowding. 
Simon \& Clement (1993) performed
simulations that added constant light 
to RR Lyrae light curves and found that $\phi_{31}$
remained unchanged. 
By taking this approach, we avoid using the RR Lyrae
$M_V-\rm{[Fe/H]}$ relation and the problems that arise
because of the uncertainty of  its
slope.

In \S \ref{sec-identM3like}, we discuss our
sample selection. Then in 
\S \ref{sec-crowd}, we use 
the CCC period-amplitude relation
to identify the crowded stars and 
perform simulations to ascertain the nature of their unresolved 
companions. 
In \S \ref{sec-extinction}, we use 
CCC's M3
period-temperature relation to calculate the temperature for each 
LMC star so that the interstellar extinction 
and corrected magnitude $V_0$ can be derived
and in \S \ref{sec-tilt}, we consider the effects of the
LMC geometry on the apparent magnitude of the RR Lyrae variables. 
Finally in
\S \ref{sec-distance}, we derive an LMC distance.

\section{The Analysis}
\label{sec-analysis}

\subsection{Identification of M3-like variables in the LMC}
\label{sec-identM3like}

\begin{figure}
\includegraphics{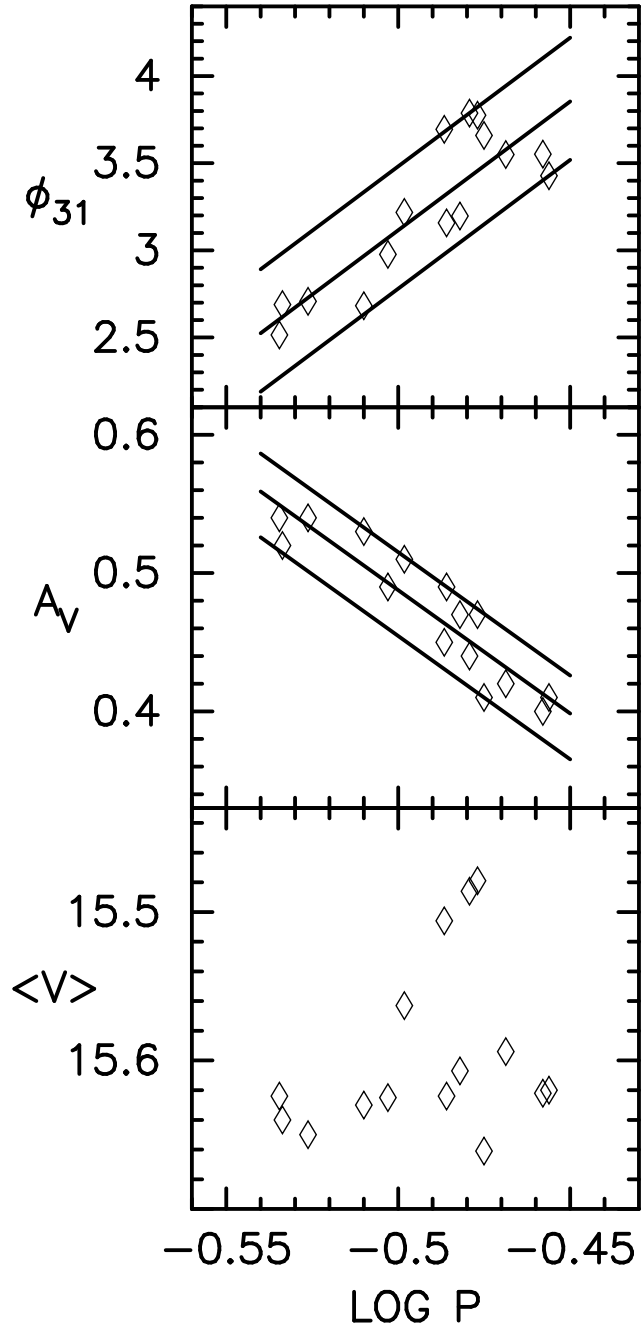}
\caption{ Plots of $\phi_{31}$, $V$ amplitude and the (intensity) mean
$V$ magnitude versus $\log P$ for the 15 M3 unevolved RR1 variables
that we include in our sample. The data
are taken from the investigation by CCC.
In the two
upper panels, the central lines represent least squares fits to
the data and the outer lines, plotted with the same slope, are
the envelope lines that encompass all of the data. 
\label{fig-M3rr1}}
\end{figure}

In their seminal study of the M3 RR Lyrae variables, CCC published
photometric and Fourier parameters for  23 variables that they
classified as type RRc (RR1). 
We use their $V$ data, but limit our study to the unevolved stars.
Thus we exclude V70, V85, V129, V170 and V177.
These five stars have longer periods for a given amplitude 
than the others and appear to have evolved off the ZAHB. 
CCC called them ``long$P$/overluminous" stars.
Another three stars, V105, V178 and V203, are excluded
because they have short periods ($P<0.29$ days)
and small amplitudes.  They do not fit into the same
period-amplitude sequence as the other RR1 variables.\footnote{Some 
of these short period variables could 
be second overtone (RR2) pulsators.
CCC plotted the Fourier parameters $\phi_{21}$
versus $A_{21}$ for V105, V178 and V203 and concluded that at least
V203 is an RR2 variable.}
 This leaves 15 `unevolved' RR1 variables in our M3 reference sample. 
In Figure \ref{fig-M3rr1}, we plot $\phi_{31}$, 
$A_V$ (the $V$ amplitude),  and
$\langle V \rangle$ 
against $\log P$ for these stars. The diagram illustrates
that $\phi_{31}$ and $A_V$ are both correlated
with period, but $\langle V \rangle$ is not. The central lines in 
the $\phi_{31}$ and
$A_V-\log P$ plots are least squares fits to the data. The outer
lines have the same slope and are envelope lines that encompass all of
the data. 
It is important that none of these 15 stars
is affected by crowding. To check this, we verified that there
was no correlation between the observed $V$ magnitude and $\Delta A_V$,
the displacement from the central line in the $A_V-\log P$ plot of
Figure \ref{fig-M3rr1}. Furthermore, the M3 finding chart published
by Bailey (1913) indicates that all of these variables are located
outside the cluster core. Therefore it seems reasonable to assume that
crowding does not affect our reference sample.

\begin{figure}
\includegraphics{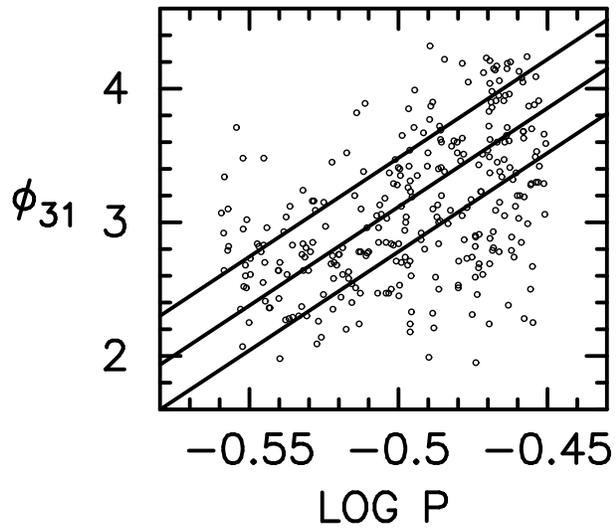}
\caption{A plot of the Fourier phase difference $\phi_{31}$ vs
$\log P$ for 330 bona fide RR1 variables in 16 MACHO fields in
the LMC. The superimposed
lines represent the $\phi_{31}- \log P$ relation for the M3
RR1 variables studied by CCC. 
(See the upper panel of Figure \ref{fig-M3rr1}.)
\label{fig-LMCperphi}}
\end{figure}

\begin{figure}
\includegraphics{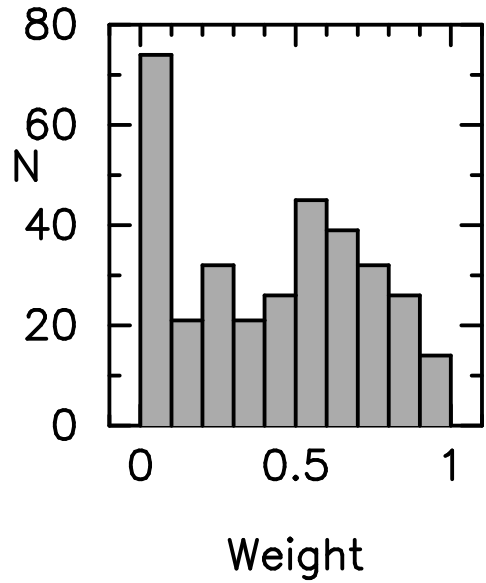}
\caption{The distribution of `weights' for the LMC
RR1 variables in  our sample. 
The weight is a measure of the probability that
a star lies between the envelope lines of Figure
\ref{fig-LMCperphi}. 
Our procedure for determining
these weights is described in
\S \ref{sec-identM3like}.
\label{fig-wthist}}
\end{figure}

Our LMC data are the MACHO data for 330 RR Lyrae
variables in 16 
fields\footnote{A chart showing the locations of
the MACHO fields is available at http://www.macho.mcmaster.ca}
near the LMC bar. The observations were obtained between 1992
and 1999. The 330 stars that we analyse were all classified as
bona fide RR1 variables by A04 who published their
photometric and Fourier parameters. 
Figure \ref{fig-LMCperphi} shows the $\phi_{31} - \log P$ plot for these
stars, with the envelope lines for M3 superimposed.
LMC variables with $\log P > -0.45$ (equivalent to $P=$ 0\day 355)
have been excluded from the plot because 
M3 RR1 variables with periods  greater than this appear 
to have evolved. 
In order to decide which of the LMC stars to include in our
sample, we used a weighting scheme that Gladders \& Yee
(2000) devised for determining whether data points belong to a linear
sequence, given their error. 
We constructed a Gaussian  distribution for 
$\phi_{31}$ of each star, using a HWHM equal to the error listed by
A04. Then we determined a `weight' by measuring
the fraction of the area under the 
Gaussian that fell between
the envelope lines of Figure \ref{fig-LMCperphi}.
Figure \ref{fig-wthist} shows 
the distribution of these weights. For
our analysis, we will include only the stars 
with the highest probablility of fitting the M3 $\phi_{31}-\log P$
plot,  i.e. the ones with weight greater than $0.5$.

\subsection{The Crowded Stars}
\label{sec-crowd}

\subsubsection{Correction for crowding} 
\label{sec-corcrowd}

\begin{figure}
\includegraphics{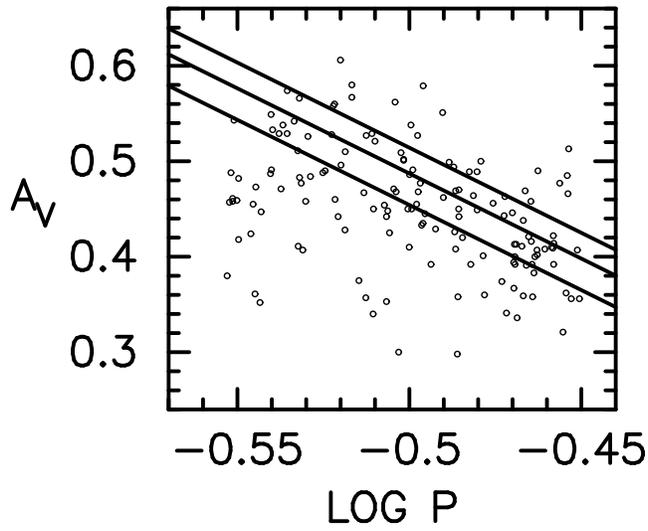}
\caption{The period-$V$ amplitude relation for the M3-like RR1
variables in 16 MACHO fields in the LMC. Only
the 147 stars with weight greater than $0.5$ in Figure \ref{fig-wthist}
are included. The superimposed lines represent the
period-amplitude relation for the M3 RR1 variables (see the centre 
panel of Figure \ref{fig-M3rr1}).
\label{fig-LMCperamp}}
\end{figure}

The period-$V$ amplitude relation for the M3-like RR1
variables in the LMC is shown in Figure \ref{fig-LMCperamp} with
the envelope lines from Figure \ref{fig-M3rr1} superimposed. 
Of the 147 stars plotted, 71 lie below the lines, 54 lie
between them and 22 are above. If our hypothesis that low
amplitudes are caused by blending is correct, we would expect
the stars below the lines to be brighter. However, before
we proceed to test this hypothesis, we need to consider
the stars with the shortest periods ($\log P< -0.54$, i.e. $P<$ 0\day 29). 
They all have low
amplitudes and could therefore
be the LMC counterparts of the M3 stars, V105, V178 and V203.
Since their low amplitudes might be intrinsic, they
are not suitable for our crowding
test. We therefore exclude all LMC variables with
$P<$ 0\day 29 from our analysis. 
This leaves 127 stars: 54 below the lines, 51 between and 22 above. 
The mean magnitudes for the stars in these three
regimes are listed in Table \ref{tbl-magpa1}. 
Each row represents a different threshold for the weights
of the stars considered. 
We will base our discussion on all of the stars with weight $> 0.5$. 

The data of Table \ref{tbl-magpa1} demonstrate that the
stars that lie below the M3 period-amplitude relation are brighter than the 
ones that lie between the lines.  
This is the result we expect if the low amplitude stars are blended. 
A t-test (for all the stars with weight $>0.5$)
indicates that the difference in $\langle V \rangle$ is highly 
significant, with a probability of only $0.0056$
that the two groups of stars are drawn from the same population.
As for the stars that lie above the M3 lines, the difference
between their mean
$\langle V \rangle$ and that of the stars between
the lines is not significant.
Two M3 counterparts to these stars, V85 and V177,
are displaced with respect to the other RR1 variables
in the period-amplitude relation,
but their $\phi_{31}$ and $\langle V \rangle$ values do not
set them apart. 
CCC classified them with their long period/high luminosity group.
We therefore suggest that these high amplitude stars are evolved, 
even though they do not appear brighter than the others.

Thus we can account for 
the relative mean magnitudes of the stars below, between  and above the 
lines in Figure \ref{fig-LMCperamp}. 
In this discussion, we have not taken the effect of interstellar
extinction into account and we know that it
may vary from star to star.
However, if the average extinction among the stars in each
of the three regimes of Figure \ref{fig-LMCperamp} is the same, 
the ranking of their mean $\langle V \rangle$ values should be correct. 

Table \ref{tbl-magpa1} indicates    that the $V$ magnitudes
of 54 of 127 M3-like RR1 variables appear to be altered by crowding.
This represents approximately 40\% of the stars.

\subsubsection{Crowding Simulations}
\label{sec-crowdsim}

If the RR1 stars that lie below the envelope lines 
in the period-amplitude diagram have unresolved
companions, what are the apparent magnitudes of these companions? 
To answer this question, we performed simulations to ascertain how
the presence of an unresolved companion would affect the
observed 
magnitude and amplitude. A few examples of these simulations
are presented in Table \ref{tbl-crowdsim}.
The simulations show the change in $V$ magnitude and amplitude
when an RR1 variable with $V=19.4$ or $19.7$ mag
is blended with a star with $V=20-22$ mag.
This range of $V$ magnitudes was selected 
for the companions because
the LMC color magnitude diagram
plotted by Alcock et al. (2000a) shows 
a high density of main sequence stars with $V>20$ mag and therefore
the RR Lyrae are probably blended with stars like these.
In Table \ref{tbl-blendsim}, we show simulated data for
three stars that
are displaced from the central line in Figure \ref{fig-LMCperamp} 
by more than $0.1$ mag.  
According to the table, the `true' magnitudes of these 
RR Lyrae stars could be $V= 19.70$, $19.56$ and $19.63$
while their 
unresolved companions have $V$ magnitudes,  $21.25$, $20.28$ and 
$20.50$ respectively. 
These RR Lyrae $V$ magnitudes are typical LMC RR1 magnitudes
and the $V$ magnitudes of the companions are consistent with
LMC main sequence stars that belong to a younger population.
RR Lyrae variables belong to an old population for which the main
sequence turn-off is about $3.4$ mag fainter than the horizontal
branch. Thus older main sequence stars would be too faint to
have much of an effect.

The MACHO CM diagram also shows a high concentration of RR Lyrae
variables and horizontal branch
red clump stars with $V\sim 19.2$ mag, but we do not
expect the stars in our data set to be blended with stars this
bright.  LMC RR Lyrae
variables that have unresolved companions with $V\sim 19.2$ mag would
be brighter than 19th magnitude. Such stars are known to exist in the
MACHO database, but the  147 stars in our sample are fainter.
In an independent study of LMC RR Lyrae variables, 
Di Fabrizio et al.  (2005, hereafter referred to as 
DF05)\footnote{A major study of RR Lyrae variables near the LMC bar 
was made by 
Clementini et al. (2003, hereafter referred to as C03) and DF05. 
They made photometric observations
with the $1.54$ m Danish telescope at La Silla, Chile and
spectroscopic observations with the $3.6$ m ESO telescope and the VLT.
They discovered approximately 135 RR Lyrae
variables in two fields that overlap with parts of MACHO fields 6 and
13.}
identified five  stars that could be blended with red clump stars 
among the RR Lyrae stars in their sample. 
These five  stars had 
typical RR Lyrae periods, but  small amplitudes and bright
mean $V$ ($<19$) magnitudes.
They concluded that one of these anomalous stars was an RR0 blended
with a young main sequence star, but did not assign a definite
classification to the other four.

\subsection{The Extinction}
\label{sec-extinction}
A serious difficulty in deriving the distance to LMC stars is
that the amount of interstellar extinction is not constant. 
Schwering \& Israel (1991) estimated that the foreground extinction due
to dust in the Galaxy ranges from $E(B-V)=0.07$ to $0.17$ over the
LMC surface. Furthermore Harris et al. (1997) concluded that the
distribution of dust within the LMC itself is clumpy. Therefore it is
desirable to derive the extinction for the stars individually 
and we are in a position to do this. Since 
we have selected M3-like stars for our investigation,
we can calculate the  temperature of each star individually
and then derive its reddening.
We use CCC's M3 period-temperature
relation:
\begin{eqnarray}
A= 13.353 -1.19 \log P_0 - 4.058 \log T_{eff}.
\end{eqnarray}
CCC found that a value of $A=-1.82\pm 0.03$
gave the best fit for the unevolved variables, but it
predicted temperatures that were too low for stars that had
evolved away from the ZAHB.
Assuming that $A=-1.82$
and that
$\log P_0 = \log P_1 + 0.127$,\footnote{A typical ratio
$P_1/P_0$ for the M3 RR01 variables (Clementini et al. 2004)
is $0.746$. This is equivalent to $\Delta \log P=0.127$.}
we derive the following equation
for calculating the temperatures of the unevolved LMC stars:
\begin{eqnarray}
\log T_{eff}=3.702-0.293 \log P_1.
\end{eqnarray}
The
unreddened color $(V-R)_0$ can be computed from a relation
derived by Kov\'acs \& Walker (1999) based on the models of
Castelli, Gratton \& Kurucz (1997a):
\begin{eqnarray}
 \log T_{eff}=3.8997-0.4892(V-R)_0+0.0113 \log g + 0.013 \rm{[M/H]}
\end{eqnarray}
For this calculation, we assume that [M/H] = $-1.3$, the
value adopted by CCC for their M3 study,\footnote{We also note that 
there is a small, but non-zero metallicity dependence subsumed into CCC's
estimation of $A$.} 
and $\log g =2.93$, the mean of the $\log g$ values they calculated
for the 15 stars in our reference sample.
In order to calculate the
corrected  magnitudes $V_0$, we assume
a ratio of total to selective absorption:
\begin{eqnarray}
A_V/E(V-R)=5.35 
\end{eqnarray}
(Schlegel, Finkbeiner \& Davis 1998).

\begin{figure}
\includegraphics{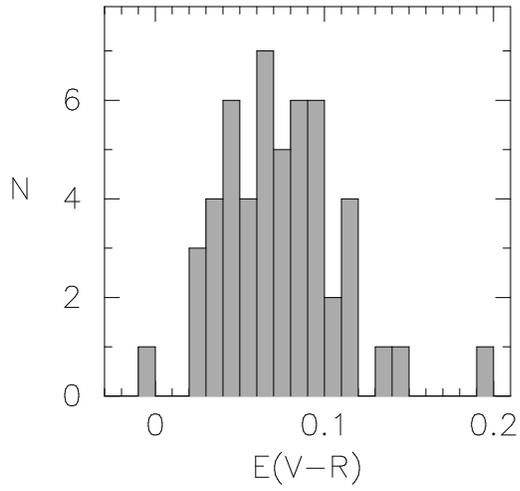}
\caption{A histogram of E(V-R) for the 51 uncrowded, unevolved M3-like 
RR1 variables in 16 MACHO fields in the LMC.
These are the stars that lie between the envelope lines of
Figure \ref{fig-LMCperamp}.
\label{fig-LMCexthist}}
\end{figure}

The $\langle V \rangle _F$ and $\langle R \rangle _F$
magnitudes,\footnote{Our $\langle V \rangle$ and
$\langle R \rangle$ values were 
derived by fitting a 6-order Fourier series of the form: 
\begin{eqnarray}
mag = A_0+\sum _{j=1,6} A_j \cos (j\omega t + \phi _j)  
                     \hskip 2mm 
\end{eqnarray}
to the $V$ and $R$ magnitudes for each star, 
where $\omega$ is ($2\pi$/period). Thus our mean magnitudes are
the $A_0$ values from the fit of equation (5) to the observational data.}
the $V$ extinction and
corrected magnitudes $V_0$ for the 51 uncrowded, unevolved
stars are listed in Table \ref{tbl-paramm3}.
Figure \ref{fig-LMCexthist} shows the distribution of $E(V-R)$ for
the stars that we consider to be uncrowded, unevolved M3-like RR1
variables. The mean is $0.073 \pm 0.02$\footnote {A04 estimated
that the error in $E(V-R)$ due to uncertainties in the temperature-color
transformation would be $\sim 0.01$ mag. In addition, there is an error
of $\sim 0.015$ which arises from the uncertainty in the value of
$A$ in equation (1).  
Combining these in quadrature leads to an error
of $0.018$ mag in $E(V-R)$ and $0.10$ in the $V$ extinction.}
which corresponds to $V$ band extinction of $0.39 \pm 0.10$.
This value can be compared with the LMC extinction that
Zaritsky et al. (2004) derived
using a different technique. They measured effective temperatures
and line-of-sight extinction for millions of individual stars
by comparing stellar atmosphere models with $U, B, V, I$ photometry.
Then they constructed extinction maps for
stars in two temperature ranges where the model
fitting between temperature  and extinction was least degenerate: 
$5500 K \le T_E < 6500 K$ (cool, older stars) and 
$12,000 K \le T_E < 45,000 K$
(hot, younger stars). They derived a mean $V$ absorption of 
$0.43$ mag for the cool stars and $0.55$ mag for the hotter stars. 
With temperatures in the range $\sim 6100$ to $7300 K$, RR Lyrae
variables are similar to their cool star group and the mean extinction
we have derived agrees with theirs to within our quoted error.
However, for the cool stars, Zaritsky et al. found a bimodal distribution 
which they attributed to the existence of a dust layer.
This bimodal structure is not evident in our data, but our  sample
is several orders of magnitude smaller than theirs.
A Shapiro-Wilk W test indicates that our data, plotted
in Figure  \ref{fig-LMCexthist}, do not deviate significantly from a
normal distribution.

We can also compare our extinction with the values derived by
C03 for their RR Lyrae investigation
since the two LMC fields they observed overlap with
MACHO fields 6 and 13. They derived $E(B-V) = 0.116\pm 0.017$ for
their field A and $0.086\pm 0.017$ for field B from the colors of
the edges of the instability strip. The  corresponding total
$V$ extinction values are $0.36\pm 0.05$ and $0.27\pm 0.05$. The
mean extinction we calculate for the 9 stars in MACHO field 6
is $0.35$ mag and for the 4 stars in field 13, it is $0.38$.
These are in good agreement with the extinction C03 derived
for their field A and a bit high compared to their field
B value, but still within the quoted errors.

We have pointed out that equation (2) predicts temperatures that
are too low for stars that have evolved away from the ZAHB. 
Thus if the stars above the lines in Figure \ref{fig-LMCperamp}
have evolved, we would expect the extinction derived from
equations (2), (3) and (4) to be underestimated and this would
lead to faint $V_0$ values.  This is exactly what we see in
Table \ref{tbl-magpa2} where we list the mean $V_0$ 
these equations predict. 
Even though their mean $\langle V \rangle$ is comparable to the mean
for the stars between the lines,
their $V_0$ is $0.08$ mag fainter. 
We conclude that this supports our hypothesis that these stars have
evolved. 
We will not include them in the sample of
unevolved M3-like RR1 variables we use to determine the LMC
distance.
For the stars below the lines of Figure \ref{fig-LMCperamp},
equation (2) should be valid for computing temperatures, but
if they are blended with other stars, their observed $(V-R)$
colors are not their true ones. Thus the $V_0$ values
we derive for these stars individually will also be in error. 
However, if the
mean extinction for these stars is comparable to the
mean extinction for the
stars between the lines, they will still appear brighter  and this is what
Table \ref{tbl-magpa2} illustrates.

\subsection{Tilt Correction and Line-of-Sight Distribution}
\label{sec-tilt}

The distribution of the  RR Lyrae population in the LMC has not been
well established. Kinematic studies by Freeman et al. (1983),
Storm et al. (1991) and Schommer et al. (1992) all indicated
that the  oldest globular clusters belong to a flattened
disk-like system with $\sigma _{RV} \sim 28 \rm{kms}^{-1}$. 
There was no evidence for the presence of a halo.
It was therefore assumed that the RR Lyrae variables must belong to
a disk population.
However, van den Bergh (2004) pointed out that the
observed radial velocities did not rule out the possibility that the globular
clusters formed in a halo. Subsequently, radial velocities
of LMC RR Lyrae variables derived by Minniti et al. (2003)
and by Borissova et al. (2006) have indicated a larger velocity distribution
($\sigma _{RV}= 50 \pm 2 \rm{km s}^{-1}$) and these authors 
argue that there is an old and metal poor halo in the LMC.
If the RR Lyrae variables belong to a halo population, they should
be distributed spherically with respect to the LMC center.
On the other hand, if they belong to a disk population,
the tilt of its plane with respect to the plane of the
sky is an effect that must be considered  when deriving the distance.

We made tilt corrections
based on recent investigations of the LMC geometry by van der Marel
\& Cioni (2001) and by Nikolaev et al. (2004).
Van der Marel \&
Cioni (2001) analysed the variations in brightness
of asymptotic 
and red giant branch stars  in near-IR color magnitude
diagrams extracted from the DENIS and 2MASS surveys.
They found a sinusoidal variation in apparent magnitude
as a function of position angle,  which they
interpreted to be the result of distance variations 
because one side of the LMC plane is closer to us than the
other.
For their analysis, they 
assumed that the LMC center is located at $\alpha _0= 82\deg 25$
and $\delta _0 = -69\deg 5$
(van der Marel 2001) and included
stars with $\rho$ in the range  $2\deg 5 - 6\deg 7$
where $\rho$ is the angular distance from the LMC center. 
They 
derived an inclination angle $i = 34\deg 7  \pm 6\deg 2$
and line-of-nodes position angle $\Theta = 122\deg 5 \pm 8 \deg 3$.
Nikolaev et al. (2004) carried out a similar analysis based
on more than 2000 MACHO Cepheids with $\rho < 4^{\circ}$.
Assuming $\alpha _0= 79 \deg 4$ and $\delta _0= -69\deg 03$,
they derived $i = 30\deg 7  \pm 1\deg 1$
and $\Theta = 151\deg 0 \pm 2 \deg 4$.

\begin{figure}
\includegraphics{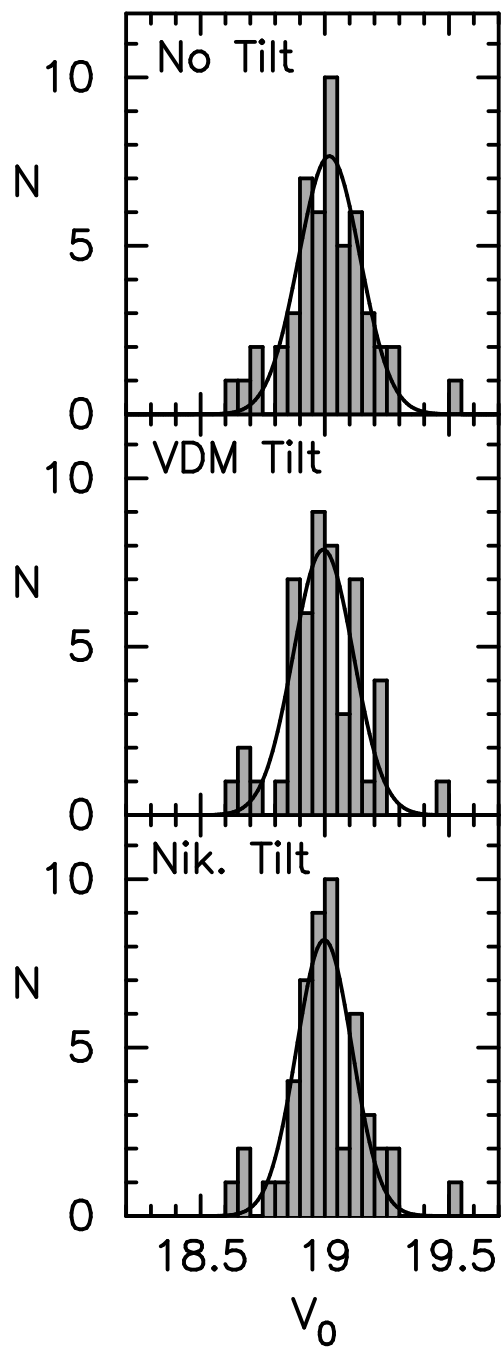}
\caption{A histogram for the LMC variables
that lie between the lines in Figure 
\ref{fig-LMCperamp}. These are the stars that
we consider to be uncrowded M3-like unevolved RR1 variables. 
Three histograms
are plotted: the first represents the data with 
no tilt correction
and the other two include the tilt corrections of van der Marel
\& Cioni (2001)
and Nikolaev et al. (2004) respectively.
The curves are Gaussian fits to the data. 
\label{fig-LMChist}}
\end{figure}

We used the equations listed in \S 2 of van der Marel \& Cioni's paper
to calculate corrected $V_0$ magnitudes for both inclinations
and they are listed in columns  (7) and (8) of
Table \ref{tbl-paramm3}.
Figure \ref{fig-LMChist} shows the distribution of
$V_0$ based on the three different assumptions
for the LMC tilt.
All three show a peak at approximately $19.0$ mag.
Some of the dispersion in $V_0$ is due to depth within the LMC, but
we expect a dispersion in absolute magnitude as well because
our M3 reference stars have a range of $0.2$ in apparent 
magnitude.\footnote{The error in $V_0$ due to the error in extinction
is $\pm 0.10$ mag. However, this is a systematic error and should not
affect the shape of the $V_0$ distributions plotted in Figure
\ref{fig-LMChist}.}
The normal parameter estimates are listed in Table \ref{tbl-normal}. 
A Shapiro-Wilk W test indicates
that none of the three deviates significantly from
 a normal distribution.

We adopt the tilt corrections
based on the viewing angles derived by Nikolaev et al. because
the stars in their sample, like ours, are all within
$4^{\circ}$ of the LMC center. The mean $V_0$ based on these
viewing angles is exactly the same as the value obtained when
no tilt correction is applied. Therefore our derived $V_0$ does not
depend on any assumption about the distribution of the LMC RR Lyrae
variables.

\section{The LMC Distance}
\label{sec-distance}

\subsection{The Apparent Magnitude of the RR1 Variables}
\label{sec-appmag}

We have adopted a mean $V_0 = 19.02$ for the 51
uncrowded, unevolved M3-like RR1 variables, based on the LMC
viewing angles derived by Nikolaev et al. (2004).
Since our mean magnitudes are the $A_0$ values from
equation (5), we need to convert them to intensity means
before we compute an LMC distance. 
A04 showed that the intensity means are  $\sim 0.01$ mag brighter
than $A_0$ so
we revise our adopted mean $V_0$ to $19.01$ mag.

To determine the precision of the $V_0$ values listed in Table
\ref{tbl-paramm3}, we need to consider
the errors in reddening and the errors
in the calibration of our photometry.
We have already pointed out in \S \ref{sec-extinction} that the error
in $E(V-R)$ is $0.018$ mag which corresponds to $0.10$ mag in
the $V$ exinction.
Our calibrated $V$ and $R$ magnitudes are from the study by A04, and
were calculated from transformation
equations derived by Alcock et al. (1999),
designated calibration version 990318.
They derived an internal precision of $\sigma _V = 0.021$ for
their $V$ magnitudes 
by comparing the results for stars in overlapping (MACHO) fields.

Another way to test our calibration is to 
compare with the results from other investigations.
Two of the 51 uncrowded, unevolved M3-like RR1 stars that we have
listed in Table \ref{tbl-paramm3} were observed by DF05.
The mean (intensity) magnitudes derived from the two studies for these
stars are listed in Table \ref{tbl-photcomp}
and they agree to within $0.01$ mag.
In our earlier paper (A04), we listed mean  magnitudes for
five additional RR1 stars that were observed by both groups and the
MACHO magnitudes were significantly brighter. 
However, in our new analysis, we have classified two of these stars
as crowded. 
Their mean magnitudes are also listed
in Table \ref{tbl-photcomp} and it is clear that the MACHO
magnitudes are brighter than the ones derived by DF05.
The remaining three
are not M3-like so we can not ascertain whether or not 
they are crowded.

DF05 also compared their $V$ magnitudes with MACHO data provided
by Alves and by Kov\'acs (Alcock et al 2003)
and found that the MACHO magnitudes
were approximately $0.04$ mag brighter on average.
DF05 concluded that the difference
occurred because the MACHO  reduction procedure did not adequately compensate 
for crowding in stars with $V>  18.25$. The MACHO collaboration
recognized the crowding
problem so A04 made crowding corrections by adding artifical
stars of known
magnitude to the the image frames and measuring $\Delta V$ the
difference between their input and recovered magnitudes. 
However, these corrections
introduce a relatively large  error of $\sim 0.10$ mag into the adopted
apparent magnitudes. 
In the present  investigation, we have dealt with this problem
by identifying 
crowded stars and removing them from our data set.

We can not determine which of the MACHO stars in the Alves and
Kov\'acs samples
were crowded; however, our sample of M3-like stars can provide some insight.
According to the data listed in Table \ref{tbl-magpa1}, 
approximately 40\% of the
M3-like variables are crowded and as  a consequence,
their average $V$ magnitude is $0.09$ mag brighter. 
If we assume that the mean MACHO magnitude for 
40\% of the stars in the data sets that Alves and Kov\'acs provided to
DF05 is
$0.09$ mag brighter (as a result of blending with main sequence stars), 
while the remaining 60\% have $V$ magnitudes
similar to the ones that DF05 derived,
we would expect the MACHO stars to be about $0.04$ mag brighter
on average. Thus if the crowded
stars could be removed from the MACHO sample, the photometry
derived in these two independent studies would be in good agreement.

Since our $V$ photometry for uncrowded stars appears to be
in good agreement with
that of DF05, we assume that the uncertainty in our calibrated
$V$ magnitudes is $0.02$ mag. 
Therefore we
adopt $V_0 = 19.01 \pm 0.10$ (extinction) $\pm 0.02$ (calibration)
for our sample of
uncrowded, unevolved M3-like variables in the LMC.
This is in good agreement with the result of C03: 
$V_0= 19.064 \pm 0.064$
at [Fe/H]$=-1.5$ on the metallicity scale of Harris (1996). 
Taking into account the fact that
the RR1 variables in our sample are M3-like, that [Fe/H] for M3
is $-1.57$ on the Harris scale and that C03 derived $\Delta M_V/\Delta
\rm{[Fe/H]}= 0.214$ for the RR Lyrae variables in the
LMC, we estimate that $V_0 = 19.05$ at [Fe/H]$=-1.57$
for the data of C03. 

\subsection{The Absolute Magnitude and Distance of the RR1 Variables}
\label{sec-absmag}

A major challenge in determining the LMC distance
is to identify a homogeneous group of stars for which
the absolute magnitude is well established. 
We deal with this problem by using the M3
distance modulus to calculate the absolute $V$
magnitude of our 15 M3 reference stars. Then we assume that
the uncrowded stars in our LMC sample have the same mean $M_V$. The M3 distance
modulus has been derived in investigations
of RR Lyrae variables by Marconi et al. (2003) 
and by Sollima et al. (2006).

Taking a theoretical approach, Marconi et al. 
applied pulsation theory
to the $BV$ observations of M3 by
Corwin \& Carney (2001) and the $K$
observations of Longmore et al. (1990). 
They compared the results from pulsation theory
with the observed edges of the instability
strip,  the observed $K$ band period-magnitude relation
and the observed relations among period-magnitude-color and
period-magnitude-amplitude. 
For their calculations, 
they used bolometric corrections and temperature-color transformations
provided by Castelli, Gratton \& Kurucz (1997a, 1997b)
and adopted a mean RR Lyrae mass of
$0.67 \msun$, based on evolutionary models of Cassisi et al. (2004).
They computed a mean distance
modulus $DM = 15.07 \pm 0.05$,\footnote{The evolutionary distance modulus that
Marconi et al. derived was about $0.08\pm 0.05$ longer than the
pulsational value. However, they 
stated that it would be shorter if element diffusion were properly 
taken into account because
the luminosity of HB models would be about $0.03-0.04$
mag fainter.}
but pointed out that if they had used the 
models of Vandenberg et al. (2000) instead, 
$DM$ would have  been $15.05$.
We adopt the distance modulus based on the Cassisi models,
DM=$15.07 \pm 0.05$.
Our 15 M3 reference stars have $V$ magnitudes ranging from 
$15.48$ to $15.68$ with mean
$\langle V \rangle =15.60$. If we apply CCC's extinction,
$E(B-V)=0.01 \pm 0.01$, this corresponds to $V_0 = 15.57$ 
which leads to $M_V=0.50 \pm 0.06$.

Sollima et al. derived an M3 distance modulus 
from the RR Lyrae period-metallicity-$K$ band
luminosity $(PL_K)$ relation that they calibrated from observations:
\begin{eqnarray}
M_K=-2.38 (\pm 0.04) \log P_F + 0.08 (\pm 0.11) \rm{[Fe/H]}
-1.05 (\pm 0.13)
\end{eqnarray}
where $M_K$ is the absolute $K$ magnitude, $P_F$ is the fundamentalized
pulsation period and [Fe/H] refers to the metallicity scale of Carretta
\& Gratton (1997). 
They derived their
coefficient of $\log P_F$ from the slope of the $K-\log P_F$
relation for 538 RR Lyrae variables in 16 globular clusters
and their [Fe/H]  coefficient  from the slope of the 
$(M_K-2.38 \log P)$ - [Fe/H] relation for
the four globular clusters in their sample that had
distance determinations based on \it Hipparcos \rm trig parallaxes for
local subdwarfs. They obtained their zero point from the $K$ magnitude
of RR Lyrae combined with the  
trig parallax that Benedict et al. (2002) measured for it from
HST astrometry. From equation (6), Sollima et al. calculated an
M3 distance modulus  of
$15.07$. Thus $M_V=0.50 \pm 0.20$.

Unfortunately, the large error in the coefficient
of [Fe/H] in equation (6) results in a large uncertainty in $M_K$. 
Therefore it seems appropriate to calculate the
M3 distance modulus directly from the absolute magnitude that
Benedict et al. derived for RR Lyrae: $M_V=0.61^{-0.11}_{+0.10}$. 
The [Fe/H] for RR Lyrae is $-1.39$ (Clementini et al. 1995), 
which is comparable to the M3 metal abundance, $-1.34$ on the scale
of Carretta \& Gratton. Furthermore,
the pulsation period of RR Lyrae, 0\day 567, and its maximum $V$
amplitude\footnote{RR Lyrae 
exhibits the
Blazhko effect, but according to Szeidl (1988), the maximum
light amplitude for a Blazhko star fits the period amplitude relation
for singly periodic variables.} 
which is $0.9$  mag according to
Smith et al. (2003) places it on the M3 period-amplitude relation.
CCC studied five RR0 stars with periods within $0.01$ days of
0\day 567  (V10, V69, V135, V137 and V142).
The mean $V$ amplitude
for these stars is $0.89$ mag and their mean $\langle V \rangle$
is $15.65 \pm 0.02$. Assuming that their mean absolute magnitude is the same as
that of RR Lyrae,
we derive a mean $M_V = 0.56\pm 0.11$ for our 15 M3 reference stars.

The Baade-Wesselink technique has not been applied to any RR Lyrae
variables in M3. However, in a review of RR Lyrae luminosities,
Cacciari (2003) reported the result of a B-W analysis for the
star RR Cet which has a metal abundance comparable to that of M3.
Cacciari et al. (2000) derived  $M_V = 0.56 \pm 0.15$ for this star
which, with a 
period of 0\day 553 and $V$ amplitude $0.98$ mag (Simon \& Teays 1982)
fits the M3 P-A relation. Six stars analysed by CCC
(V36, V40, V71, V89, V133
and V149) have periods within $0.01$ days of 0\day 553 and the  
mean $V$ amplitude and mean $\langle V \rangle$ they derived for these
stars were $0.99$ and $15.65$ respectively. By comparing the mean
magnitudes of these six
stars with those of our 15 M3 reference stars, we derive mean
$M_V= 0.52 \pm 0.15$ for the reference stars.

The mean of our four $M_V$ values is $0.52$ with a standard deviation 
$0.02$ mag. 
Combining this with the 
$V_0=19.01 \pm 0.10$ (extinction error) $\pm 0.02$ (calibration error)
that we derived for the uncrowded, unevolved
M3-like RR1 variables in our LMC sample and adding the estimated
errors in quadrature, leads to distance
modulus $\mu _ {LMC}= 18.49 \pm 0.11$. 

Our distance modulus is in good agreement with
$\mu _0= 18.48 \pm 0.08$ derived by Borissova et al.
(2004) from $K$-band photometry of 37 RR Lyrae variables
in the inner regions of the LMC. In their investigation,
they derived a mean $\langle K
\rangle = 18.20$ and assumed that the mean $K$ band absorption
$A_K= 0.05$ mag. 
By following the procedure described by Bono et al. (2001, 2003),
they calculated $M_K=-0.332$ at $\log P = -0.30$.
Their adopted  absolute $K$ magnitude was $0.85$ mag
brighter than our adopted $M_V$. This is consistent with
the apparent magnitudes we have derived; their mean
$K_0$ ($18.15$) is $0.86$ mag brighter than our $V_0$, $19.01$.
Our adopted $\mu _0$ is also comparable to the
value $(18.48)$ obtained by McNamara et al. (2007) in their recent
analysis of $\delta$ Scuti stars. 

By identifying the crowded
stars and removing them from the sample, we have avoided
using the crowding corrections that introduced an additional uncertainty
of $0.10$ mag to the distance modulus we derived in our 
earlier study (A04). The main source of error in this investigation
is the error in estimating the $V$ extinction.

\section{Summary}
\label{sec-summary}

We have devised a new method for identifying crowded
RR1 variable stars in the LMC, based on simulations that show that
stars with unresolved companions have low amplitudes 
for their periods.
Given that many LMC RR Lyrae variables have properties similar to the 
ones in the Galactic globular cluster M3, we used the
M3 $\phi_{31}-\log P$ relation to identify the M3-like unevolved
RR1 variables in our LMC sample.
The Fourier phase parameter $\phi _{31}$ is useful for selecting
a homogeneous sample because it is not affected by crowding.

When the M3-like variables were plotted on the M3 period-amplitude
diagram, we found that the mean $V$ magnitude of the LMC stars
with low amplitudes was $0.09$ mag brighter than the mean for the 
stars that fit the M3 period-amplitude relation. 
Four of the stars in our sample were observed 
in the study of LMC RR Lyrae variables by DF05. 
Comparing the photometry from the two studies, we found that our
$V$ magnitudes for the two stars considered to be uncrowded
agreed to within $0.01$ mag, while the MACHO $V$
magnitudes for the two stars we considered to be crowded were 
more than $0.05$ mag brighter. From this, we conclude that
our method for identifying crowded RR Lyrae variables is effective.
It could prove to be useful for identifying crowded RR Lyrae
variables in other local group galaxies.

We used the M3 period-temperature relation for
unevolved RR Lyrae variables to determine the temperature
and reddening for each of the uncrowded RR1 variables in our
sample. After making corrections for the tilt of
the LMC, we derived a mean $V_0$ magnitude of $19.01 \pm 0.10$
(extinction) $\pm 0.02$ (calibration).
Then to estimate the absolute magnitude, we
used the M3 distance modulus and the trig parallax of
RR Lyrae to derive the mean absolute magnitude
of the unevolved RR1 variables in our M3 reference sample. This
turned out to be $M_V= 0.52 \pm 0.02$. 

Finally, we 
derived an LMC distance modulus $\mu _{LMC} = 18.49 \pm 0.11$
which is in good agreement with the results of other recent studies
and
with $18.5$ mag, the  value  employed by the Hubble Space Telescope's
key project for measuring the Hubble constant (Freedman et al.
2001).

\acknowledgements

We thank David Clement and 
Doug Welch for their helpful comments during the preparation of this
manuscript.  We also express our gratitude to our referee,
Bruce Carney, who made several suggestions that have improved
the paper.
In addition, financial support from 
Science and Engineering Research Canada (NSERC) is gratefully
acknowledged.

\clearpage

\clearpage

\clearpage

\clearpage

\clearpage

\clearpage

\clearpage

\begin{deluxetable}{lcccccc}
\tablecaption{Mean $\langle V \rangle$ and Location in the
Period-Amplitude  Diagram
\label{tbl-magpa1}}
\tablewidth{0pt}
\tablehead{
\colhead{Weight} &
\colhead{Mean $\langle V \rangle$} &
\colhead{N} &
\colhead{Mean $\langle V \rangle$} &
\colhead{N} &
\colhead{Mean $\langle V \rangle$} &
\colhead{N} \\

\colhead{} &
\colhead{(below)} &
\colhead{} &
\colhead{(between)} &
\colhead{} &
\colhead{(above)} &
\colhead{} \\

\colhead{(1)} &
\colhead{(2)} &
\colhead{(3)} &
\colhead{(4)} &
\colhead{(5)} &
\colhead{(6)} &
\colhead{(7)} 
}
\startdata

Wt. $\ge 0.5$ & $19.31\pm 0.02$ & 54 & $19.40 \pm 0.02$ & 51 &
                $19.43\pm 0.04$ & 22 \\
Wt. $\ge 0.6$ & $19.30\pm 0.02$ & 43 & $19.40 \pm 0.02$ & 33 &
                $19.45\pm 0.05$ & 16 \\
Wt. $\ge 0.7$ & $19.28\pm 0.03$ & 31 & $19.38 \pm 0.02$ & 25 &
                $19.37\pm 0.05$ & 8 \\
Wt. $\ge 0.8$ & $19.32\pm 0.03$ & 17 & $19.35 \pm 0.03$ & 13 &
                $19.45\pm 0.06$ & 3 \\
Wt. $\ge 0.9$ & $19.31\pm 0.05$ & 9 & $19.39 \pm 0.03$ & 6 &
                $19.40\pm 0.06$ & 2 \\
\enddata
\tablecomments{The terms `below', `between' and `above' refer
to the location relative to the lines in Figure \ref{fig-LMCperamp}.
The weights represent the probability that the stars lie between
the  envelope lines in Figure \ref{fig-LMCperphi}. Stars with
$P<$0\day 29  are not included. 
}
\end{deluxetable}

\begin{deluxetable}{lcccc}
\tablecaption{Crowding Simulations 
\label{tbl-crowdsim}}
\tablewidth{0pt}
\tablehead{
\colhead{$V$} &
\colhead{Amplitude} &
\colhead{$V$} &
\colhead{$V$} &
\colhead{Amplitude} \\

\colhead{(RR)} &
\colhead{(RR)} &
\colhead{(Companion)} &
\colhead{(Observed)} &
\colhead{(Observed)} \\

\colhead{(1)} &
\colhead{(2)} &
\colhead{(3)} &
\colhead{(4)} &
\colhead{(5)} 
}
\startdata
19.4 & 0.45 & 20.0  & 18.90 & 0.29 \\
19.4 & 0.45 & 20.5  & 19.06 & 0.33 \\
19.4 & 0.45 & 21.0  & 19.17 & 0.37 \\
19.4 & 0.45 & 21.5  & 19.25 & 0.39 \\
19.4 & 0.45 & 22.0  & 19.30 & 0.41 \\
19.7 & 0.45 & 20.0  & 19.08 & 0.26 \\
19.7 & 0.45 & 20.5  & 19.27 & 0.30 \\
19.7 & 0.45 & 21.0  & 19.41 & 0.35 \\
19.7 & 0.45 & 21.5  & 19.51 & 0.38 \\
19.7 & 0.45 & 22.0  & 19.57 & 0.40 \\
\enddata
\tablecomments{The simulated magnitudes listed in columns 
(1) and (3) combine to produce the magnitudes
listed in column (4). Thus if an RR Lyrae variable with $V=19.4$
has an unresolved companion with $V=20.0$, its observed
magnitude will be $V=18.90$.
As a result of the blending, its original amplitude ($0.45$ mag)
will be reduced to $0.29$ mag.
}
\end{deluxetable}

\begin{deluxetable}{lccccc}
\tablecaption{Sample Magnitudes for Three Blended
Stars
\label{tbl-blendsim}}
\tablewidth{0pt}
\tablehead{
\colhead{Star} &
\colhead{$\log P$} &
\colhead{Amplitude} &
\colhead{$\langle V \rangle$} &
\colhead{$\langle V \rangle$} &
\colhead{Amplitude} \\

\colhead{} &
\colhead{} &
\colhead{(observed)} &
\colhead{(observed)} &
\colhead{(RR + companion)} &
\colhead{(true)} \\

\colhead{(1)} &
\colhead{(2)} &
\colhead{(3)} &
\colhead{(4)} &
\colhead{(5)} &
\colhead{(6)} 
}
\startdata
80.6475.3548 & -0.486 & 0.36 & 19.46  & 19.70+21.25 & 0.45  \\
80.6708.6879 & -0.503 & 0.30 & 19.10  & 19.56+20.28 & 0.46  \\
81.8398.799  & -0.486 & 0.30 & 19.22  & 19.63+20.50 & 0.44  \\
\enddata
\tablecomments{The simulated magnitudes listed in column (5)
combine to produce the observed magnitudes of column (4). 
For example, our simulation implies that star $80.6475.3548$ 
is really an RR Lyrae with $V=19.70$ blended with an unresolved companion
whose $V=21.25$ mag. As a 
result of the blending, the original $V$ amplitudes listed in
column (6) are reduced to the observed amplitudes listed in column (3).
}
\end{deluxetable}

\begin{deluxetable}{lcccccccc}
\tablecaption{Parameters of the Uncrowded, Unevolved  M3-like RR1 Variables
\label{tbl-paramm3}}
\tablewidth{0pt}
\tablehead{
\colhead{Star} &
\colhead{Period} &
\colhead{$\langle V \rangle _F$} &
\colhead{$\langle R \rangle _F$} &
\colhead{Ext$(V)$} &
\colhead{$V_0$} &
\colhead{$V_0$(vdM)} &
\colhead{$V_0$(Nik)} &
\colhead{Weight} \\

\colhead{(1)} &
\colhead{(2)} &
\colhead{(3)} &
\colhead{(4)} &
\colhead{(5)} &
\colhead{(6)} &
\colhead{(7)} &
\colhead{(8)} &
\colhead{(9)} 
}
\startdata

2.4789.946   & 0.326910 & 19.23 & 19.05 & 0.18 & 19.05 
  & 19.05 & 19.04 & 0.89 \\
2.5150.896   & 0.337123 & 19.21 & 19.01 & 0.24 & 18.97 
  & 18.97 & 18.96 & 0.59 \\
2.5151.982   & 0.344523 & 19.33 & 19.10 & 0.37 & 18.96 
  & 18.96 & 18.95 & 0.63 \\
2.5269.422   & 0.318123 & 19.52 & 19.29 & 0.48 & 19.04 
  & 19.03 & 19.03 & 0.90 \\
2.5511.772   & 0.302923 & 19.33 & 19.17 & 0.18 & 19.15 
  & 19.15 & 19.15 & 0.93 \\
2.5633.1369  & 0.293763 & 19.30 & 19.13 & 0.27 & 19.03 
  & 19.03 & 19.03 & 0.71 \\
\enddata
\tablecomments{
The $V_0$ values are corrected for extinction; 
the $V_0$(vdM) and $V_0$(Nik) values are corrected
for tilt as well, according to the LMC viewing angles derived by
van der Marel \& Cioni (2001) and by Nikolaev et al. (2004)
respectively.
The weight is the probability that the star lies between
the  envelope lines of Figure \ref{fig-LMCperphi}.
Table \ref{tbl-paramm3} is presented in its entirety in the
electronic edition of the Astronomical Journal. A portion is shown here
for guidance regarding its form and content.}
\end{deluxetable}

\begin{deluxetable}{lccc}
\tablecaption{Mean $V_0$ for Stars 
with Wt. $> 0.5$
\label{tbl-magpa2}}
\tablewidth{0pt}
\tablehead{
\colhead{Location in } &
\colhead{Mean $V_0$} &
\colhead{} &
\colhead{N} \\

\colhead{Fig \ref{fig-LMCperamp}} &
\colhead{} &
\colhead{} &
\colhead{} 
}
\startdata
Below the lines & $18.88 \pm 0.02$ && 54 \\
Between the lines & $19.02 \pm 0.02$  && 51 \\
Above the lines & $19.10 \pm 0.05$  && 22 \\
\enddata
\tablecomments{The $V_0$ values have been derived under the
assumption that the stars' temperatures can be computed
from equation (2) and that their observed $(V-R)$ colors
are correct. These two assumptions are valid for the
stars that lie between the lines, but not for the others.
This is discussed in the final paragraph of
\S \ref{sec-extinction}.
}
\end{deluxetable}

\begin{deluxetable}{lccc}
\tablecaption{Fitted Normal Parameter Estimates
for Fig \ref{fig-LMChist}
\label{tbl-normal}}
\tablewidth{0pt}
\tablehead{
\colhead{Tilt correction} &
\colhead{$\mu$} &
\colhead{$\sigma$} &
\colhead{N} 
}
\startdata

None & 19.02 & 0.16 & 51 \\
van der Marel & 19.00  & 0.15 & 51 \\
Nikolaev & 19.02   & 0.16 & 51 \\
\enddata
\end{deluxetable}

\begin{deluxetable}{lcccc}
\tablecaption{Comparison with the DF05 Photometry
\label{tbl-photcomp}}
\tablewidth{0pt}
\tablehead{
\colhead{ID} &
\colhead{Crowded?} &
\colhead{$\langle V \rangle _{int}$} &
\colhead{ID} &
\colhead{$\langle V \rangle _{int}$} \\

\colhead{(MACHO)} &
\colhead{(MACHO)} &
\colhead{(MACHO)} &
\colhead{(DF05)} &
\colhead{(DF05)} 
}
\startdata
6.7054.710 & no & 19.45 & 7864 & 19.46 \\
13.5838.667 & no & 19.39 & 7648 & 19.38 \\
6.6689.563 & yes & 19.23 & 2249 & 19.37 \\
13.6079.604 & yes & 19.24 & 4749 & 19.31 \\
\enddata
\end{deluxetable}

\end{document}